# A New Security Boundary of Component Differentially Challenged XOR PUFs Against Machine Learning Modeling Attacks

Gaoxiang Li[1*], Khalid T. Mursi[2*], Ahmad O. Aseeri[3],
Mohammed S. Alkatheiri[2] and Yu Zhuang[1]

[1]Department of Computer Science, Texas Tech University, TX 79409, USA
[2]Department of Cybersecurity, College of Computer Science and Engineering,
University of Jeddah, Jeddah 21959, Saudi Arabia
[3]Department of Computer Science, College of Computer Engineering and Sciences,
Prince Sattam Bin Abdulaziz University, Al-Kharj 11942, Saudi Arabia

*Abstract*

*Physical Unclonable Functions (PUFs) are promising security primitives for resource-constrained network nodes. The XOR Arbiter PUF (XOR PUF or XPUF) is an intensively studied PUF invented to improve the security of the Arbiter PUF, probably the most lightweight delay-based PUF. Recently, highly powerful machine learning attack methods were discovered and were able to easily break large-sized XPUFs, which were highly secure against earlier machine learning attack methods. Component-differentially-challenged XPUFs (CDC-XPUFs) are XPUFs with different component PUFs receiving different challenges. Studies showed they were much more secure against machine learning attacks than the conventional XPUFs, whose component PUFs receive the same challenge. But these studies were all based on earlier machine learning attack methods, and hence it is not clear if CDC-XPUFs can remain secure under the recently discovered powerful attack methods. In this paper, the two current most powerful two machine learning methods for attacking XPUFs are adapted by fine-tuning the parameters of the two methods for CDC-XPUFs. Attack experiments using both simulated PUF data and silicon data generated from PUFs implemented on field-programmable gate array (FPGA) were carried out, and the experimental results showed that some previously secure CDC-XPUFs of certain circuit parameter values are no longer secure under the adapted new attack methods, while many more CDC-XPUFs of other circuit parameter values remain secure. Thus, our experimental attack study has re-defined the boundary between the secure region and the insecure region of the PUF circuit parameter space, providing PUF manufacturers and IoT security application developers with valuable information in choosing PUFs with secure parameter values.*

## 1. Introduction

### 1.1. Overview and Motivation

The Internet-of-Things (IoT) has broad and deep penetration in business activities and people's daily lives, forming diverse groups of networks. For many of them, security is of critical importance for ensuring the integrity of communications [23,39]. But many network nodes are resource-constrained, like sensors and IoT devices, and cannot accommodate conventional cryptographic protocols, which are not lightweight, as pointed out by [15], [38]. Physical Unclonable Functions leverage small physical variations of a small number of transistors to produce responses unique to individual circuits and hence are highly lightweight and not reproducible [12,13,17,21]. Thus, PUFs are good candidates as hardware primitives for implementing security protocols for resource-constrained network nodes.

However, before adopting PUFs as a trusted security function, they must be examined to identify all possible security vulnerabilities, such as vulnerabilities to machine learning (ML) modeling attacks. In machine learning modeling attacks, the attacker eavesdrops on on-air packets between the communications of IoTs and collects enough plain challenge-response pairs (CRP) to build a model. Then, the attacker inputs the collected challenges, as features, and responses, as class labels, into an ML model targeted at having a learned model for future response prediction.

The arbiter PUF [20] is probably the most lightweight strong PUF which admits exponentially many challenge-response pairs (CRPs) but suffers from vulnerability to ML attacks. XOR PUFs [31] were proposed to improve security against ML attacks. A security study [28] showed that small-sized XOR PUF succumbed to the LR attack method, a method tailor-designed according to the circuit structure of the XOR PUF. Since 2018, studies using neural networks found that some larger XOR PUFs, including 64-bit 8-XOR PUFs and 128-bit 7-XOR PUFs [5, 30], were also not secure. And more recently, further developed attack methods [10,35] gained substantially improved modeling power and were able to break XOR PUFs of sizes that have withstood all previous ML attacks even with magnitudes fewer CRPs.

To improve the attack resistance while keeping the PUF architecture lightweight, it is a natural idea to use different challenges for different components. An XOR-PUF with different challenges for different components, also known as a component-differentially challenged XOR-PUF (CDC-XPUF), has two advantages over other alternative PUF designs. First, because CDC-XPUFs share the same structural design as XOR-PUFs, there is no need to overly increase circuit complexity or implementation resources to implement CDC-XPUFs. Second, CDC-XPUFs have exceptional modeling attack resistance. Previous research [26,33,38] established that CDC-XPUFs can achieve significant resistance against ML modeling attacks under the same XOR-PUF architecture, and there is no successful attack report for the modeling attack on 64-stage CDC-XPUFs with five components. However, these studies on the modeling attack resistance of CDC-XPUFs are only limited to previous ML attack methods, and the security performance of CDC-XPUFs has not been evaluated by more powerful approaches. Therefore, with the arrival of powerful ML methods [10, 35] for attacking XPUFs, it is natural to question whether this potential secure PUF candidate will remain secure, and in this paper, we are rightly focused on investigating this question.

In this paper, the attack study of CDC-XPUFs is carried out using the most popular neural network-based method [35] and the adapted LR method [37], both of which use parameter values adapted for the CDC-XPUF architecture and have not been applied to CDC-XPUFs before. Our experimental study shows that the LR-based method outperforms the most popular NN method and significantly reduces the number of CRPs and training time while significantly increasing attack success rates compared with previous research. CDC-XPUFs with 6 or fewer component PUFs and 128-bit CDC-XPUFs with 5 or fewer component PUFs can be broken using our adapted LR-based method, whereas no previous studies were able to break 64-bit CDC-XPUFs with more than 4 component PUFs [33], and CDC-XPUFs with 7 or more component PUFs remain well beyond the power of the two ML attack methods. Although our method does not fully break the CDC-XPUFs for all conceivable sizes and complexity yet, it reveals a vulnerability in CDC-XPUFs and the PUF designer should revise the current protocol to reduce the number of allowed CRPs in the lifespan to prevent attackers from gathering enough CRPs.

## 1.2. Organization of this Paper

The remainder of this paper is organized as follows: Section 2 lists related work about the investigation of PUF security to ML modeling attacks. And Section 3 gives a general overview of PUFs. Section 4 illustrates the evaluation tools used to break CDC-XPUFs. The experimental setup, including silicon and simulated CRP generation, and experimental evaluation results are presented in Section 5. Finally, concluding remarks are given in section 6.

## 2. RELATED WORK

PUFs are promising security primitives for resource-constrained IoT devices. Therefore, identifying all possible security vulnerabilities is one of the most important tasks in PUF studies, such as vulnerabilities to machine learning (ML) modeling attacks [4,5,22,30] and reliability-based attacks [1,9]. In machine learning modeling attacks, the attacker eavesdrops on on-air packets between IoTs and collects enough plain challenge-response pairs (CRPs) to build a model. Then, the attacker inputs the collected challenges, as features, and responses, as class labels, into an ML model targeted at having a learned model for future response prediction. In reliability-based attacks, the attacker applies pre-set challenges to the PUF with an open interface and collects specific CRPs. Furthermore, CRPs obtained through freely queried can be easily used to break PUFs by utilizing the reliability information of these CRPs.

Many efforts have been invested in the research of PUF security, especially in the vulnerability to ML modelling attacks. The work of Majzoobi et al. [22] was the first to investigate the ML modeling attack resistance of PUFs, showing that arbiter PUFs are vulnerable to ML modeling attacks.

Moreover, the first comprehensive PUF attack study was conducted by Ruhrmairet al. [28,29], which employed multiple attack methods including Support Vector Machines (SVMs), Evolution Strategies (ES), and logistic regression (LR) with resilient propagation (RProp) to break PUFs. This work further improved the performance of existing attacking methods and revealed that small-sized XOR PUF still succumbed to the LR attack method.

In 2012, a two-layer neural network [10] was used to attack 2-XOR PUFs, with four neurons for a hidden layer and one neuron for the other layer. In 2018, studies by Asseri et al. and Santikellur et al., which utilized neural networks-based attack methods with three ReLU-based hidden layers and one sigmoid output layer, found that some larger XOR PUFs, including 64-bit 8-XOR PUFs and 128-bit 7-XOR PUFs [5, 30], are also not secure.

And more recently, further developed attack methods [33,35], which applied the hyperbolic tangent function (tanh) as the activation function of NN based model, gained substantially improved modeling power and were able to break XOR PUFs of sizes that have withstood all previous ML attacks even with magnitudes fewer CRPs. Moreover, this NN-based modeling attack method has been called the most powerful attacking method and has been widely usedfor PUF-breaking tasks and the evaluation of the PUF modeling attack resistance. Though it is believed that tanh sometimes suffers from the vanishing gradient problem (which made it less popular after the introduction of ReLU) in other research fields, it seems that during the training of the networks for attacking PUFs, tanh works substantially better than ReLU.

All the previous research showed that the current popular modeling attacks based on the logistic regression modelor the artificial neural network model are highly effective at breaking XOR-PUFs. Therefore, current XOR-PUFs with popularly implemented sizes are no longer secure. To ensure security against ML attacks, increasing the PUF circuit complexity by adding additional circuit designs (stages or components) is the simplest method. However, as the number of components or stages grows, so does the cost of hardware and operational power, making them unsuitable for resource-constrained IoT devices. As a result, there is rising concern that XOR-PUF designs may suffer as a result of PUF designers being forced to provide security at the expense of dramatically increased overhead.

The aforementioned studies suggest that XOR-PUFs with component-differential challenges (CDC-XPUFs) likely have even higher resistance against machine learning attacks than XOR PUFs with the same challenge for all components. However, the analysis of CDC-XPUF's modeling attack resistance has been very limited so far. The study of Wisiol et al. [33] on pseudorandom sub-challenges PUFs, which are the same as CDC-XPUFs, showed that the existing machine learning-based attack method for 64-bit CDC-XPUFs with four sub-challenges can attain an attack success rate lower than 90% even if using 1 million CRPs and can attain probably only a 20% success rate if using only 100 thousand CRPs.

Due to the tremendously high dimensional search space for distinct multi-dimensional inputs, the difficulty and complexity of breaking CDC-XPUFs are much higher than breaking XOR-PUFs. Therefore, our work is motivated to evaluate the modeling attack resistance of CDC-XPUFs with the current best-attacking methods, which have not been applied to CDC-XPUFs before. Different from previous studies, the attacking methods and their parameter values are modified and adapted for the CDC-XPUFs architecture to evaluate their performance for modeling attack resistance. To the best of our knowledge, this is the first study that breaks CDC-XPUFs with more than four components. And our method can even break 64-bit CDC-XPUFs with six components and 128-bit CDC-XPUFs with five components.

Although there are many alternative APUF variants and many new PUF designs proposed, such as Lightweight Secure PUFs [22], FF-PUFs [13,20], and Interpose PUF[27], to the best of our knowledge, they are still vulnerable to ML modeling attacks [7,30,32,35,36]. In this paper, we only focus on the most widely studied XOR PUFandits variant, CDC-XPUFs.

In addition, according to Yu et al. [38], CDC-XPUFs are more vulnerable to reliability attacks [9] than traditional XOR-PUFs. But this reliability attack can be prevented by applying a lockdown protocol to block the open interface. Although our improved lightweight CDC-XPUFs architecture will improve the reliability by reducing the complexity, it is not the purpose of this paper to investigate resistance to reliability attacks. In this paper, we will only focus on the machine learning modeling attack and assume that the lockdown authentication protocol is in place.

## 3. BACKGROUND INFORMATION ON PUFS

In order to clarify technical discussions in later sections, the mechanisms of the arbiter PUF, XOR-PUF, and CDC-XPUF will be briefly described in this section.

### 3.1. The arbiter PUFs

Figure 1 shows a simple case of an arbiter PUF. An $n$-bit arbiter PUF is made up of n stages, each with two multiplexers (MUXs). When giving a rising signal, the signal enters the arbiter PUF from stage one and splits into two signals. The two signals are routed through gates at each stage, and the propagation paths are determined by the challenge bit to the multiplexers at each stage. Finally, two signals reach the D flip-flop, which acts as an arbiter to determine whether the signal on the top path or the signal on the lower path arrives first. If the top path signal arrives first, the D flip-flop returns 1; otherwise, it returns 0.

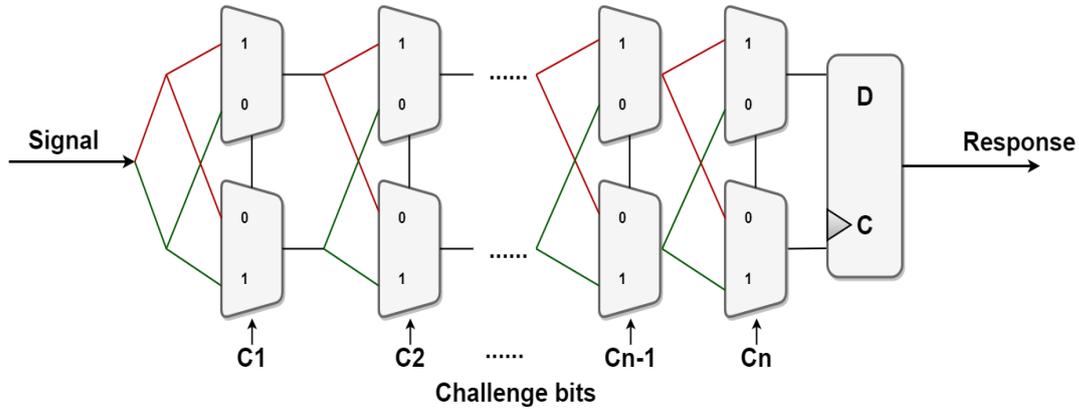

Figure 1. An arbiter PUF with n bits of challenge

### 3.2. The XOR arbiter PUFs

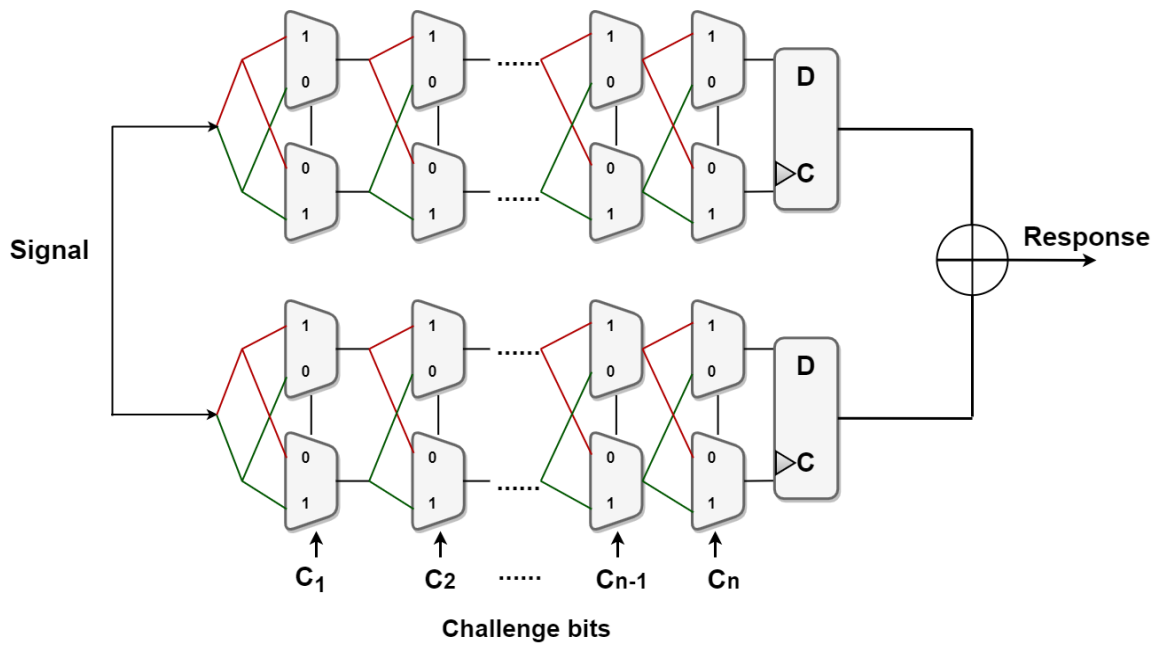

Figure 2. An XOR-PUF with 2 sub-streams and n bits of each stream

Due to arbiter PUFs' weak resistance to ML modeling attacks, a new PUF was proposed in [31] which increased a non-linear XOR gate to multiple arbiter PUFs to produce the final response. This type of PUF is known as the XOR arbiter PUF. Figure 2 illustrates a simple case of *n*-bit 3-XOR-PUF. An *n*-XOR-PUF is made up of *n* component arbiter PUFs (also known as streams or sub-challenge) in which the responses of all *n* component arbiter PUFs are XORed at the XOR gate to produce one single bit response. It is worth noting that all component arbiter PUFs in an XOR-PUF are fed the same challenge bits.

Studies in [28], [29] show that XOR-PUFs could attain higher modeling attack resistance than arbiter PUFs. When equipped with a lockdown authentication protocol [38] to eliminate open-access interface, for XOR-PUFs with 64 stages and more than 9 component arbiter PUFs, all modeling attacks developed so far were not able to crack within the limited number (100 million) of available CRPs [3-5,11,28,29,32]. Nevertheless, for 64-bit XOR-PUFs with 9 or fewer component arbiter PUFs, the most recent studies [35], [37] showed that there are attack methods that can crack and predict the responses of such PUFs with a prediction accuracy of around 98%. However, extending the number of streams and challenge stages will raise the cost and power consumption of a PUF, which is an important issue for resource-constrained IoT devices. Also, the expanding number of streams will lower the reliability of PUFs and increase the risk of reliability side-channel attacks [8,9].

### 3.3. The CDC-XPUFs

CDC-XPUF and XOR-PUF share the same architecture, which includes different multiple arbiter PUF components and XOR gates. The only difference between CDC-XPUF and XOR-PUF is that CDC-XPUF's each different component arbiter PUF receives different challenge inputs, while the XOR-PUF receives the same challenges for all its component arbiter PUFs.

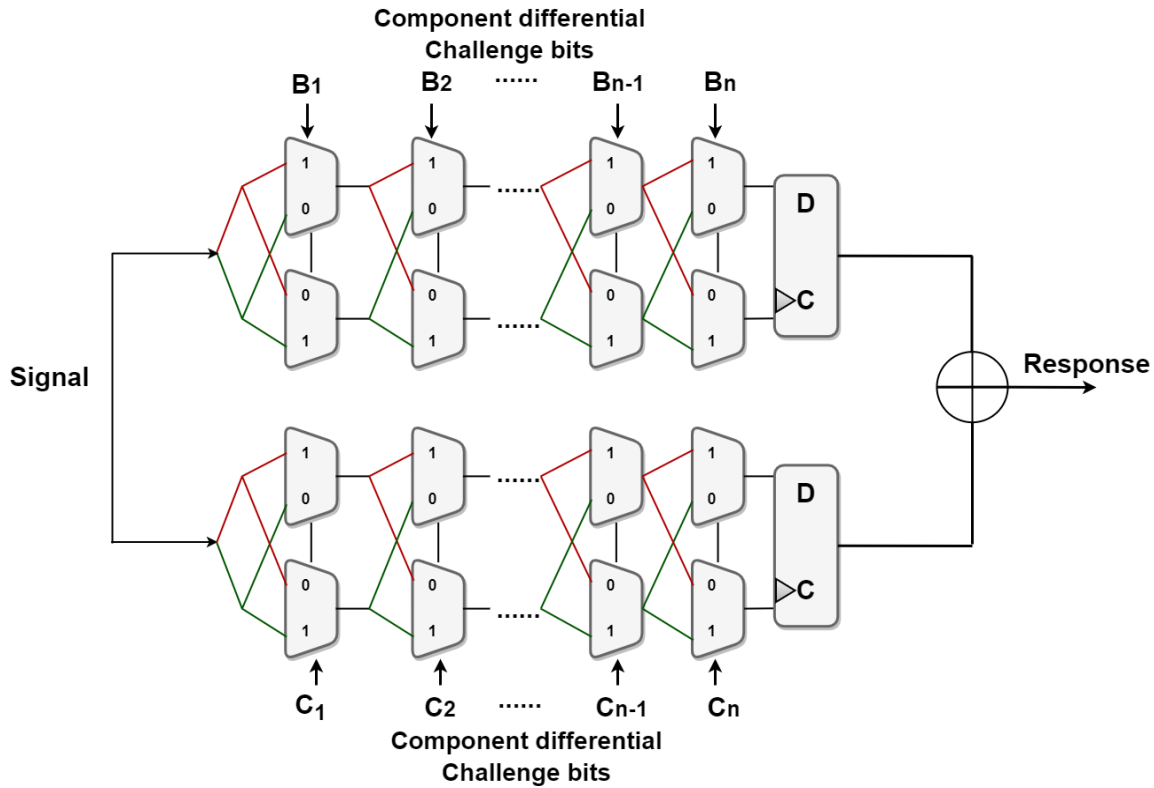

Figure 3. A CDC-XPUF with two component-differential challenges and n bits of each challenge

Studies [25,26,33,38] show that applying different challenges to different components of an XOR-PUF can decrease the vulnerability of the PUF against ML modeling attacks. Existing ML attack methods for 64-bit CDC-XPUFs with four components can attain a success rate lower than 90% even if using more than one million CRPs. All the previous experimental results show that the CDC-XPUF with four or more components is unbreakable or too expensive to break with existing attack methods. Therefore, these CDC-XPUFs can be considered potentially good candidates in terms of security performance.

## 4. ATTACKING THE CDC-XPUF

A Logistic Regression model [19] with RProp was first used to break Arbiter PUFs and XPUFs in [22,28]. And the LR-based method successfully could break Arbiter PUFs and XOR-PUFs with a small number of components [29]. Following that, the NN-based model with ReLU activation layers was used to break XOR-PUFs [5], achieving excellent results and breaking up to 8 XOR-PUFs. When breaking XOR-PUFs, the most recent study [35] proposed a NN model with the hyperbolic tangent function (tanh) as the activation function, which demonstrated significantly higher learning power than ReLU-based neural networks. This NN method significantly reduces the number of CRP required and could break 64-bit XOR-PUFs with ten components.

However, the analysis of CDC-XPUF's modeling attack resistance has been very limited so far. Related studies [33], [38] were all based on earlier machine learning attack methods, and hence it is not clear if CDC-XPUFs can remain secure under the recently discovered powerful attack methods, which have not been applied to CDC-XPUFs before. Previous studies by Wisiol et al. [33] only investigated an improved LR-based machine learning attack method for 64-bit CDC-XPUFs with four sub-challenges, and this method could only achieve a success rate of less than 90% even when using 1 million CRPs and probably as low as 20% when using only 100 thousand CRPs. To the best of our knowledge, this is the only available quantitative investigation, and no more reports are available till now.

Table 1. Parameters of the NN attack method for $k$-XOR-PUFs and CDC-$k$-XPUFs.

| Parameters | Description |
|---|---|
| Optimizing Method | ADAM [18] |
| Output Activation Function | Sigmoid |
| Learning Rate | Adaptive |
| No. Neurons in Each Layer | Layer 1 = Components × Stages<br>Layer 2 = Components × Stages/ 2<br>Layer 3 = Components × Stages/ 2<br>Layer 4 = Components × Stages |
| Loss Function | Binary cross entropy |
| Batch Size | $10^{k-1}$ |
| Kernel Initializer | Random Normal |
| Early Stopping | True, when validation accuracy is 98% |

Once a more powerful modeling attack model has been proposed and reduced the required number of CRP to break a targeted PUF instance, the security boundary of the targeted PUF needs to be updated to withstand the attacks. Therefore, the two current most powerful machine learning methods for attacking XOR-PUFs are adapted, by fine-tuning the parameters of the two methods for CDC-XPUFs, to examine the ML modeling attack resistance of CDC-XPUFs. A four-layer NN method [35], as well as the problem-tailored LR-based method based on Ruhrmari's model [29], are used to examine ML modeling attack resistance of CDC-XPUFs.

Due to the unsuitability of the multi-dimensional different challenges input into CDC-XPUFs, traditional machine learning methods such as logistic regression are inefficient in the extremely large search space and are incapable of learning the information contained within the multi-dimensional challenges. As a result, we are motivated to modify traditional logistic regression for the multi-dimensional challenges inputted in the CDC-XPUFs. To simulate the mechanism of CDC-XPUFs, the parameters of the adapted LR model are trained separately for each sub-challenge input. The output from each sub-challenge will be used as the input for the subsequent process. Figure 4 depicts an overview of the LR-based method.

The NN-based method we used in this paper is the tanh-based neural network with four hidden layers. The Tanh function is used as a hidden layer activation function, and the sigmoid function is used as an output activation function. The number of neurons in each layer is decided by the number of components and the number of stages of the tested PUF instance.

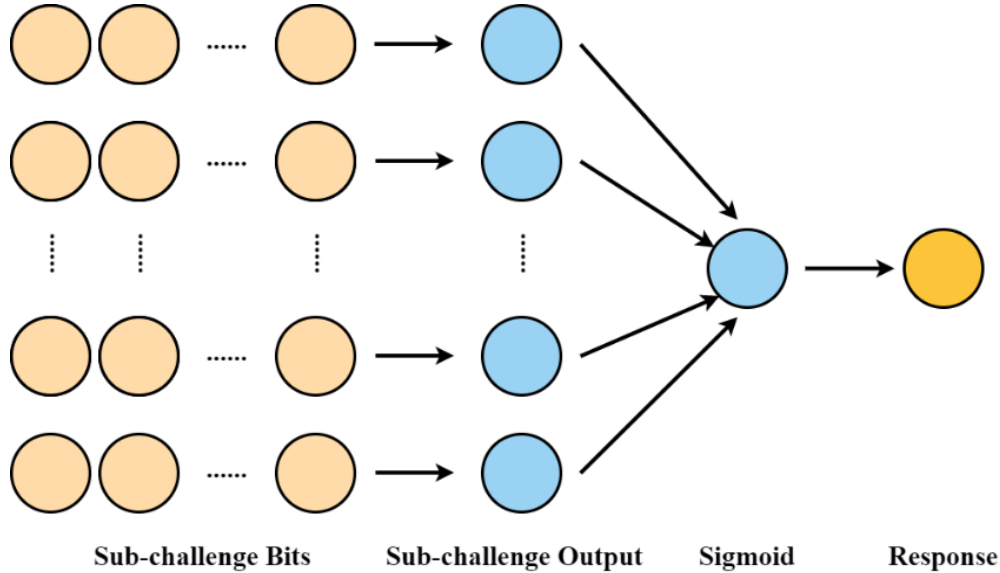

Figure 4. An overview of adapted LR-based method for attacking CDC-XPUFs with component-differential-challenges (Parameters in each sub-challenge are trained separately)

The resistance to ML modeling attacks is evaluated in particular by training an attack model with a training CRP set and then predicting responses to new challenges. The prediction rate is calculated by percentage, which means that for a 100% prediction rate, our models could correctly predict all the given challenges in the testing set. To evaluate the evaluation tools utilized in the experiment, ML modeling attacks on traditional XOR-XPUFs with the NN method and the LR-based method are also applied to verify the performance of our evaluators.

Table 2. Parameters of the LR-based attack method for $k$-XOR-PUFs and CDC-$k$-XPUFs.

| Parameters | Description |
|---|---|
| Optimizing Method | ADAM |
| Output Activation Function | Sigmoid |
| Base Learning Rate | 0.01 |
| Loss Function | Binary cross entropy |
| Batch Size | $10^{k-1}$ |
| Weight Initializer | Random Normal |
| Early Stopping | True, with 5 patience |

For the convenience of reading, the parameters of the NN attack method are listed in Table 1 and the parameters for the LR-based method are listed in Table 2. The codes implement the adapted LR-based method and the NN method will be made available for reproducible study by peer researchers when the paper is published.

## 5. EXPERIMENTAL STUDIES OF CDC-XPUFS

### 5.1. Generating CRP from Silicon and Simulator

In our experiment, two methods are used to generate the CRPs. The first is a simulator that is based on the Pypuf library [34], while the second is based on PUF instances that are implemented on silicon FPGAs.

#### 5.1.1. Generating from simulator

The simulated CRP dataset generation is based on the Pypuf simulator library and CRP generator [34]. An XOR-PUFs generator by Pypuf is modified to build the CDC-XPUF simulator used in this experiment. For each number of components set, 10 different simulated 64-bit and 10 different simulated 128-bit CDC-XPUF and XOR-PUF instances are generated with 100 million CRP. The generated PUF instances are all from the normal distribution, with a mean of 0 and a standard deviation of 1, and no noise value was added.

#### 5.1.2. Generating from Silicon

In our experiments, ten different 64-bit CDC-XPUFs and XOR-PUFs are programmed on Xilinx Artix-7 FPGA [16] [17] [24], which includes a configurable MicroBlaze CPU. The experiments are conducted on three different chips and 10 million CRPs on each instance are generated. Due to the speed and capacity constraints of our FPGA devices, PUF instances with 128-bit are not implemented and the maximum number of CRPs generated from FPGA devices is limited to 10 million. VHSIC Hardware Description Language (VHDL) is used to construct the CDC-XPUFs on Xilinx Vivado 15.4 HL design edition. The CDC-XPUFs placement in the hardware is done horizontally using Tool Command Language (TCL). Also, the Xilinx SDK is used to pass random challenges to the CDC-XPUFs and receive the corresponding response to each challenge. For generating the CRPs, AXI General Purpose Input/Output (GPIO) interfaces are used as follows: 1 GPIO for submitting the initial traveling signals, 1 for receiving the output response, and the leftovers for feeding the CDC-XPUF with the generated challenges. Our challenges are generated using the Pseudo-Random Number Generator (PRNG) as follows:

$$C_{n+1} = (a * C_n + g) \bmod m$$

where C is the sequence of the generated random number, a is a multiplier, g is a given constant, and m is 2K where K is the number of stages. To speed up the data transfer between PUFs and the computer, AXI Universal Asynchronous Receiver Transmitter (UART) was used with a bud rate equal to 230,400 bits/second. Finally, the Tera Term, which is a terminal emulator program, is used for printing and saving the device output.

Furthermore, the voltage is set to 2.0 W, the junction temperature reported by the Xilinx Artix-7 FPGA is 26℃ and the Thermal margin is 59.0℃ (12.3W) for all PUF instances. The voltage setting and temperature are controlled to the same level to the minimum impact of an external factor.

### 5.2. Experiment Setup

The experiments employ an 80-20 training-testing split, with 1% CRP from the training set used for validation. All the code is implemented in Python 3.7 using the TensorFlow and Keras ML libraries [2], [14]. Furthermore, the maximum number of CRP generated from the hardware used for experiments is 10 million, and the maximum number of CRP generated from simulators is 100 million. With the CRPs ready, we carried out tests of the attack on a high-performance computing cluster with 128 AMD EPYCTM 7702 cores and a memory capacity of 512 GB.

In each attack, the number of CRPs used in the attack starts small and gradually increases until having reached a size (the size listed in the column "Training Size" in the table), which results in a 90% attack success rate for all 10 PUF instances or a failure with 100 million CRPs. And only attacks with a testing accuracy of greater than 90% are considered successful. Moreover, due to the running time restriction of the high-performance computing cluster, the maximum training time for each attack is 48 hours. Experiments are stopped when the testing cannot converge with the maximum number of available CRP within the maximum training time.

The codes developed for the experimental study, including codes implementing the CDC-XPUFs simulator based on the Pypuf library and codes implementing two different attacking methods, will be made available for reproducible study by peer researchers when the paper is published.

### 5.3. Experimental Results

The experimental results of the ML modeling attack on CDC-XPUFs with simulator-generated CRP are listed in Tables 3 and 4 and the experimental results on FPGA-generated CRP are listed in Table 5. The results of the ML modeling attack on XOR-PUFs are also added to verify the performance of our evaluation tools.

The "Security Evaluator" column indicates which attacking method was used in this experiment for each testing result row, Security Evaluator "LR" indicates the adapted LR-based method, and Security Evaluator "NN" indicates the neural network-based method with four hidden layers and tanh as activation function. Moreover, only testing accuracy higher than 90% is considered a successful attack in the column " Success Rate " and only the accuracy of a successful attack is counted to calculate the "Average Accuracy".

The results show that, first and foremost, on both different CRP datasets, the NN attacking evaluation tools succeed in breaking most normal-sized tested XOR-PUF instances within a given number of CRPs. In addition, the LR-based evaluation tools could crack most XOR-PUF instances with small XOR-size when given 10M silicon CRP. Both methods could show solid performance for breaking XOR-PUFs. Therefore, the experimental results on breaking XOR-PUFs prove that these two evaluation tools are strong enough to examine the security vulnerability of PUFs against ML attacks.

Table 3. Experimental results on simulator generated 64-bit CRP dataset

| Number of Stages | PUF Type | Security Evaluator | Training Size | Average Accuracy | Training Time | Success Rate |
|---|---|---|---|---|---|---|
| 64 bits | 6 XOR-PUF | LR | 2.5m | 98% | 2 min | 90% |
| | | NN | 1.2m | 99% | 3 min | 100% |
| | 7 XOR-PUF | LR | 10m | 98% | 30 min | 90% |
| | | NN | 2m | 99% | 3 min | 100% |
| | 8 XOR-PUF | LR | 40m | 96% | 1 hr | 90% |
| | | NN | 10m | 99% | 5 min | 100% |
| | 9 XOR-PUF | LR | 100m | No convergence | 48 hrs | 0% |
| | | NN | 40m | 99% | 10 min | 90% |
| | CDC-3-XPUF | LR | 6k | 96% | 10 sec | 100% |

|  |  | NN | 80k | 99% | 1 min | 100% |
|---|---|---|---|---|---|---|
|  | CDC-4-XPUF | LR | 80k | 97% | 1 min | 100% |
|  |  | NN | 6m | 98% | 15 min | 90% |
|  | CDC-5-XPUF | LR | 4.5m | 96% | 30 min | 90% |
|  |  | NN | 100m | No convergence | 48 hrs | 0% |
|  | CDC-6-XPUF | LR | 100m | 95% | 20 hrs | 80% |
|  |  | NN | 100m | No convergence | 48 hrs | 0% |

For the experimental result on attacking 64-bitCDC-XPUFs, the required number of CRP (80k only) to break 64-bit CDC-XPUFs with 4 components is much smaller than the 10M required CRP in the earlier report by Wisiol et al. [33]. And the number of CRP required to break 64-bit CDC-5-XPUFs is only 4.5 million, which is still far less than the number required in previous research for CDC-4-XPUFs. With 100 million CRP, even 64-bit CDC-6-XPUFs can be broken by this adapted LR-based method. For the experimental on breaking 128-bit CDC-XPUFs, our method could break 128-bit CDC-3-XPUFs using only 50k CRP and could break128-bit CDC-4-XPUFs using 450k CRPs. Moreover, the required number of CRP to break 128-bit CDC-5-XPUFs is 40m.

Table 4. Experimental results on simulator generated 128-bit CRP dataset

| Number of Stages | PUF Type | Security Evaluator | Training Size | Average Accuracy | Training Time | Success Rate |
|---|---|---|---|---|---|---|
| 128 bits | 5 XOR-PUF | LR | 2m | 95% | 3 min | 100% |
|  |  | NN | 700k | 98% | 2 min | 100% |
|  | 6 XOR-PUF | LR | 35m | 97% | 40 min | 90% |
|  |  | NN | 10m | 98% | 3 min | 100% |
|  | 7 XOR-PUF | LR | 40m | No convergence | 48 hrs | 0% |
|  |  | NN | 20m | 98% | 3 min | 100% |
|  | 8 XOR-PUF | LR | - | - | - | - |
|  |  | NN | 80m | No convergence | 48 hrs | 0% |
|  | CDC-3-XPUF | LR | 50k | 96% | 10 sec | 100% |
|  |  | NN | 400k | 96% | 5 min | 100% |
|  | CDC-4-XPUF | LR | 450k | 97% | 2 min | 100% |
|  |  | NN | 30m | 98% | 40 min | 80% |
|  | CDC-5-XPUF | LR | 40m | 97% | 2 hrs | 50% |
|  |  | NN | 100m | No convergence | 48 hrs | 0% |
|  | CDC-6-XPUF | LR | 100m | No convergence | 48 hrs | 0% |
|  |  | NN | - | - | - | - |

A previous study by Yu et al. [38] showed the existence of a PUF system-level instantiation that is exponentially difficult to learn. So that due to the execution time and memory capacity restrictions on our tested high-performance computing clusters, the attack for 64-bit CDC-7-XPUFs and 128-bit CDC-6-XPUFs failed at 100M CRPs, and CDC-XPUFs with more sub-challenges are not eligible to be tested because the memory capacity of the estimated required CRPs exceeds the limit of the given hardware too much.

To the best of our knowledge, this is the first time to break 64-bit CDC-XPUFs with more than 4 components, allowing us to gain a better understanding of the modeling attack security boundary of CDC-XPUFs. CDC-XPUFs were almost unbreakable or too expensive to break before this study. For the time being, some previously secure CDC-XPUFs of certain circuit parameter values are no longer secure under the adapted new attack methods while many more CDC-XPUFs of other circuit parameter values remain secure. The authentication control protocol must be updated to reduce the number of allowed CRPs to ensure the safety of CDC-XPUFs in real-world applications. Otherwise, some previously secure CDC-XPUFs can be broken by this adapted LR-based attacking method once enough CRPs are eavesdropped by attackers, which is a significantly smaller number than in previous work.

In addition, the adapted LR-based method achieves significantly higher attacking performance than the most popular NN attacking method. Due to the unsuitability of the multi-dimensional different challenges input into CDC-XPUFs, traditional LR or NN methods are not efficient for the tremendously high-dimensional search space. Also, ML models must learn prior knowledge within the multidimensional challenges of CDC-XPUFs on their own, making cracking CDC-XPUFs even more difficult. This is the reason our adapted problem-tailored LR-based attacking method outperforms the general NN method in attacking CDC-XPUFs. Thus, this adapted LR-based method could be a method of evaluating the security vulnerability of CDC-XPUFs to ML modeling attacks.

## 6. CONCLUSION

In this paper, an investigation of the security of the CDC-XPUFs against the most powerful ML attack methods with problem-adapted parameter values was conducted to evaluate the security performance of CDC-XPUFs. When compared to results reported earlier, our study has found vulnerability of the CDC-XPUF with PUF circuit parameter values previously was not found to be insecure. In particular, our study has broken 64-bit CDC-6-XPUFs using around 100 million simulated CRPs and 64-bit CDC 5-XPUFs with 4.5 million simulated CRPs or 2.5 million silicon CRPs. In addition, our study also has broken 128-bit CDC-5-XPUFs using 40 million simulated CRPs. Both of them were beyond the modeling power of any earlier ML attack methods. This method also could break 64-bit CDC-4-XPUFs using only around 80 thousand CRPs, magnitudes fewer than the CRPs used in an earlier study. On the other side, this study also reveals that the security of CDC-XPUFs increases sharply as the number of component PUF increases, and 64-bit CDC XPUFs with 7 components are completely beyond the power of the two ML attack methods, which is also good news for IoT security community that there are many CDC-XPUFs remain secure, at least all CDC XPUFs with 64 bits or longer challenges and 7 or more component PUFs are secure against the most powerful ML attack methods developed so far. Therefore, our experimental attack study has re-defined the boundary between the secure region and the insecure region of the PUF circuit parameter space, providing PUF manufacturers and IoT security application developers with valuable information for the protocol of current CDC-XPUF-based applications to mitigate potential risks.

Table 5. Experimental results on FPGA generated CRP dataset

| Number of Stages | PUF Type | Security Evaluator | Training Size | Average Accuracy | Training Time | Success Rate |
|---|---|---|---|---|---|---|
| 64 bits | 6 XOR-PUF | LR | 2m | 95% | 3 min | 90% |
| | | NN | 0.3m | 96% | 30 sec | 100% |
| | 7 XOR-PUF | LR | 8m | 95% | 5 min | 90% |
| | | NN | 0.5m | 95% | 1 min | 100% |
| | 8 XOR-PUF | LR | 10m | 95% | 10 min | 90% |
| | | NN | 2m | 96% | 2 min | 100% |
| | 9 XOR-PUF | LR | 10m | No convergence | 48 hrs | 0% |
| | | NN | 4m | 96% | 20 min | 90% |
| | CDC-3-XPUF | LR | 5k | 96% | 10 sec | 100% |

|  |  | NN | 80k | 97% | 1 min | 100% |
|---|---|---|---|---|---|---|
| CDC-4-XPUF |  | LR | 80k | 96% | 1 min | 100% |
|  |  | NN | 0.5m | 95% | 5 min | 90% |
| CDC-5-XPUF |  | LR | 2.5m | 96% | 15 min | 90% |
|  |  | NN | 10m | No convergence | 48 hrs | 0% |
| CDC-6-XPUF |  | LR | 10m | No convergence | 48 hrs | 0% |
|  |  | NN | 10m | No convergence | 48 hrs | 0% |

## CONFLICTS OF INTEREST

The authors declare no conflict of interest.

## ACKNOWLEDGMENTS

We thank the anonymous reviewers whose comments/suggestions helped improve and clarify this manuscript. Also, the authors acknowledge the High-Performance Computing Center (HPCC) at Texas Tech University for providing computational resources that have contributed to the research results reported in this paper. URL: http://www.hpcc.ttu.edu

## AUTHORS

**Gaoxiang Li**: Gaoxiang Li received a B.S. degree in Computer Science and Technology from Shandong Normal University, Jinan, China, in 2017 and an M.S. degree in Computer Science from Auburn University, Auburn, USA, in 2020. Heis currently working toward a Ph.D. degree in Computer Science with the Department of Computer Science, Texas Tech University, Lubbock, USA. His research interests include machine learning, IoT security, and physically unclonable functions.

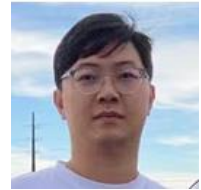

**Khalid T. Mursi**: Dr. Khalid T. Mursi received his B.S. from King Abdulaziz University (KAU), Jeddah, in 2011, in Information Technology, and his M.S. and Ph.D. from Texas Tech University (TTU), Lubbock, TX, USA, in 2018 and 2021, in Computer Science. He is currently an assistant professor at the Department of Cybersecurity – University of Jeddah. His research interests include IoT security, Machine Learning attacks, and Arabic sentiment analysis.

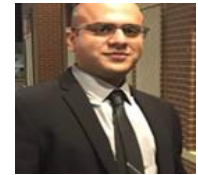

**Ahmad O. Aseeri**: Dr. Ahmad O. Aseeri received a bachelor's degree in computing from King Saud University, Saudi Arabia, in 2007, a Master of Science (M.Sc.) in Computer Science from the University of Wisconsin-Milwaukee, United States, in 2014, and a Doctor of Philosophy (Ph.D.) in Computer Science from Texas Tech University, United States, in 2018. He is currently an Assistant Professor at the Department of Computer Science, College of Computer Engineering and Sciences, Prince Sattam Bin Abdulaziz University, Saudi Arabia. He is a research collaborator at the Cyber-Physical System lab at the University of Jeddah, Saudi Arabia. Dr. Aseeri's primary research lies in the field of Artificial Intelligence with direct applications to resource-constraint IoT devices and Physical Unclonable Functions. He also has research works in data mining and modeling with application to clustering techniques, including K-means and bisecting memory-aware K-means for big data. Further, he has research interests in the deep learning methods applied to signal processing and Time Series Analysis.

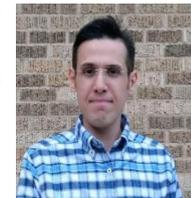

**Mohammed Saeed Alkatheiri**: Mohammed Saeed Alkatheiri received a bachelor's degree in computer science from King Abdulaziz University, Saudi Arabia, a master's degree in communication network security from Beihang University, China, and a Ph.D. degree in computer science from Texas Tech University, USA, through a full scholarship. He is currently an Associate Professor with the Department of Cybersecurity, College of Computer Science and Engineering, University of Jeddah, Saudi Arabia, and also an advisor at the Ministry of Education. His current research interests focus on the area of cybersecurity as cryptography, digital authentication, machine learning and pattern recognition, security in resource-constraint devices, and technological innovation management. He was a Researcher with the Center of Excellence in Information Assurance, King Saud University, Saudi Arabia. He served as a consultant for national projects and joined the Prince Muqrin Chair for Information Security Technology (PMC) along with government departments on National Information Security Strategy project as a security consultant.

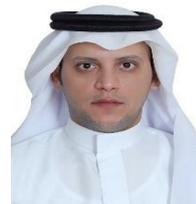

**Yu Zhuang**: Dr. Yu Zhuang received his Ph.D. in Computer Science and Ph.D. in Mathematics both in 2000 at Louisiana State University. He was a visiting assistant professor at the computer science department of Illinois Institute of Technology from April to July of 2001and has been with the Texas Tech computer science department since September 2001. Dr. Zhuang's research interests include IoT security, high-dimensional data modelling, mining, and high-performance scientific computing.

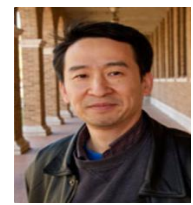